# A Comparison of Performance Measures via Online Search[*]

Joan Boyar    Kim S. Larsen    Abyayananda Maiti

University of Southern Denmark
Odense, Denmark

{joan,kslarsen,abyaym}@imada.sdu.dk

## Abstract

Though competitive analysis has been a very useful performance measure for the quality of online algorithms, it is recognized that it sometimes fails to distinguish between algorithms of different quality in practice. A number of alternative measures have been proposed, but, with a few exceptions, these have generally been applied only to the online problem they were developed in connection with. Recently, a systematic study of performance measures for online algorithms was initiated [Boyar, Irani, Larsen: Eleventh International Algorithms and Data Structures Symposium 2009], first focusing on a simple server problem. We continue this work by studying a fundamentally different online problem, online search, and the Reservation Price Policies in particular. The purpose of this line of work is to learn more about the applicability of various performance measures in different situations and the properties that the different measures emphasize. We investigate the following analysis techniques: Competitive, Relative Worst Order, Bijective, Average, Relative Interval, Random Order, and Max/Max. In addition to drawing conclusions on this work, we also investigate the measures' sensitivity to integral vs. real-valued domains, and as a part of this work, generalize some of the known performance measures. Finally, we have established the first optimality proof for Relative Interval Analysis.

---

[*]Partially supported by the Danish Council for Independent Research.

# 1 Introduction

An optimization problem is *online* if input is revealed to an algorithm one piece at a time and the algorithm has to commit to the part of the solution involving the current piece before seeing the rest of the input [4]. The first and most well-known analysis technique for determining the quality of online algorithms is *competitive analysis* [13, 18, 15]. The competitive ratio expresses the asymptotic ratio of the performance of an online algorithm compared to an optimal offline algorithm with unlimited computational power. Though this works well in many contexts, researchers realized from the beginning [18] that this "unfair" comparison would sometimes make it impossible to distinguish between online algorithms of quite different quality in practice.

In recent years, researchers have considered alternative methods for comparisons of online algorithms, some of which compare algorithms directly, as opposed to computing independent ratios in a comparison to an offline algorithm. See the references below and [9] for a fairly recent survey. Most of the new methods have been designed with one particular online problem in mind, trying to fix problems with competitive analysis for that particular problem. Not that much is known about the strengths and weaknesses of these alternatives in comparison with each other. In [7], a systematic study of performance measures was initiated by fixing a (simple) online problem and applying a collection of performance measures. Partial conclusions were obtained in demonstrating which measures focus on greediness as an algorithmic quality. It was also observed that some measures could not distinguish between certain pairs of algorithms where the one performed at least as well as the other on every sequence.

We continue this systematic study here by investigating a fundamentally different problem which has not yet been studied as an online problem other than with competitive analysis, the *online search problem* [12, 11]. Online search is a very simple online (profit) maximization problem. Prices, between the minimum price of $m$ and the maximum price of $M$, arrive online one at a time, and each time a price is revealed, the algorithm can decide to accept that price and terminate or decide to wait. The length of the input sequence is not known to the algorithm in advance, but is revealed only when the last price is given, and the algorithm must accept that price, if it has not accepted one earlier.

This simple model of a searching problem has enormous importance due to its simplicity and its application in the much more complex problems of lowest or highest price searching in various real-world applications in the fields of Economics and Finance [17]. The online search problem is very similar to that of the one-way trading problem [8, 11, 12]. In fact, one-way trading can be seen as randomized searching.

Greediness, as explored in the server problem [7], has to do with focus on which choice to make when advancing from one partial solution to another, closer to the end result. The online search problem does not have partial solutions, so the findings obtained in this paper are complementary to the results obtained in [7].

Our primary study is of the class of Reservation Price Policy (RPP) algorithms [11, 12]. This is a parameterized class, where the behavior of $\mathcal{R}_p$ is to accept the first price greater than or equal to the so-called reservation price $p$.

As a "sanity check" to confirm that the measures "work" at all on this problem, we also define $\mathcal{R}_p^2$, which accepts the *second* price greater than or equal to $p$, and investigate its relationship to $\mathcal{R}_p$. Whereas $\mathcal{R}_p$ "decides what it wants and takes it when it sees it", $\mathcal{R}_p^2$ "knows what it wants, but does not take it until the second time it sees it". One would expect $\mathcal{R}_p$ to be the better algorithm. With the exception of Max/Max Analysis, all the measures "pass this test" and favor $\mathcal{R}_p$, though significant redefinition was necessary for Relative Interval Analysis.

Having that the measures pass this test, we proceed to the more interesting task of comparing the different quality measures on RPP algorithms with different parameters. We have considered having an integral interval of possible prices between $m$ and $M$ as well as a real-valued scenario. For the most part, the results are similar. The following discussion is assuming a real-valued scenario, allowing us to state the results better typographically, without rounding.

For the performance measures below, note that since profit is a constant (between $m$ and $M$), independent of the sequence length, for measures of an asymptotic nature, we use the strict version since asymptotic results (allowing an additive constant) would deem all algorithms optimal (ratio one compared with an optimal algorithm—up to the additive constant). In each section of the paper, we give the precise definition of the measures used.

We find that Competitive Analysis and Random Order Analysis favor $\mathcal{R}_{\sqrt{mM}}$, the reason being that they focus on limiting the worst case ratio compared to an optimal algorithm, independent of input length. Relative Interval Analysis favors $\mathcal{R}_{\frac{m+M}{2}}$, similarly limiting the worst case difference, as opposed to ratio. Average Analysis favors $\mathcal{R}_M$. This is basically due to focusing on the limit, i.e., when input sequences become long enough, any event will occur eventually. In Bijective Analysis, basically all algorithms are incomparable. Finally Relative Worst Order Analysis deems the algorithms incomparable, but gives indication that $\mathcal{R}_{\sqrt{mM}}$ is the best algorithm.

In addition to these findings, this paper contains the first optimality result for Relative Interval Analysis, where we prove that no $\mathcal{R}_p$ algorithm can be better than $\mathcal{R}_{\frac{m+M}{2}}$. For Relative Worst Order Analysis, we refine the discussion of which algorithm is best by introducing the concept of "domination", which seems to be interesting for classes of parameterized algorithms. A limited use of this concept, without naming it, appeared when comparing variants of Lazy Double Coverage for the server problem in [7].

Finally, we have investigated the sensitivity of the different measures with regards to the choice of integral vs. real-valued domains, and most of the measures seem very stable in this regard. For Bijective Analysis, the situation changes from basically all algorithms being incomparable to all algorithms being equivalent. Average analysis is inapplicable for a real-valued interval, but a generalization, which we call Expected Analysis, can be applied, giving similar results to what Average Analysis gives for integral values.

Note that our problem is a profit maximization problem. Thus, for those analysis methods which have previously only been defined for cost minimization problems, we have presented profit maximization versions. Both Relative Interval Analysis and the Max/Max ratio have previously been defined with limits which were inappropriate for a problem such as online search where the maximum value achievable by an algorithm does not grow with the number of requests. This also led to considering alternative definitions which are more appropriate for this type of scenario.

The rest of this paper is organized as follows. Section 2 defines the notation used and each subsequent section treats one of the measures described above.

## 2  Problem Preliminaries

Unless otherwise stated, we assume that the prices are integral and drawn from some integral interval $[m, M]$ with $0 < m \leq M$. In any time step, any value from this closed interval can be drawn as a price, and there will be $N = M - m + 1$ possible prices. This assumption is made for the sake of consistency; some methods of analysis are uninteresting for real-valued intervals; see Section 4, for example. Also, this assumption is compatible with the real-world problems of online search as the set of prices is generally finite (the market decides on an agreed-upon number of digits after the decimal point).

We denote the length of the price sequence by $n$. Denote by $\mathcal{I}_n$ the set of all input sequences of length $n$. Thus, the total number of possible input sequences of length $n$ is $N^n$. For an online algorithm $\mathcal{A}$ and an input sequence $I$, let $\mathcal{A}(I)$ be the profit gained by $\mathcal{A}$ on $I$, i.e., the price chosen. In some analyses (for example in Relative Worst Order Analysis), we need to permute the input sequences. We always use $\sigma$ as a permutation and denote the permuted sequence by $\sigma(I)$.

We use $\mathcal{R}_p$ to denote the RPP algorithm with reservation price $p$, i.e., the algorithm that accepts the first price of at least $p$ and accepts the last price of the input sequence if it has not accepted another price before that. We let $\mathcal{R}_p^2$ denote the similar algorithm that accepts a price of at least $p$ the second time such a price is seen.

Some of the analysis methods compare two online algorithms directly (for example Bijective Analysis) and some methods (for example Competitive Analysis) compare the online algorithms with a hypothetical optimal offline algorithm which receives the input in its entirety in advance and has unlimited computational power in determining a solution. We denote this optimal algorithm by $OPT$ and the profit gained by it from an input sequence $I$ as $OPT(I)$, which is the maximum price in that sequence.

To denote the relative performance of two online algorithms $\mathcal{A}$ and $\mathcal{B}$ according to an analysis method, $x$, we use the following notation. If $\mathcal{B}$ is better than $\mathcal{A}$, then we write $\mathcal{A} \prec_x \mathcal{B}$, and if $\mathcal{B}$ is no worse than $\mathcal{A}$, this is denoted by $\mathcal{A} \preceq_x \mathcal{B}$. If the measure deems the algorithms equivalent, then this is denoted by $\mathcal{A} \equiv_x \mathcal{B}$. Usually, we merely define either $\prec_x$ or $\preceq_x$ and the other relations follow in the standard way from that.

To streamline the presentation of results, we always assume that $n \geq 2$. This is because if the input sequence contains a single price, i.e., $n = 1$, the first and only price is also the last price. By the definition of the problem, the online player has to select that single price, and all online search algorithms are equivalent.

The core of the paper is concerned with the comparison of $\mathcal{R}_p$ and $\mathcal{R}_q$ for $p \neq q$. In order not to have to state this every time, we always assume that $m \leq p < q \leq M$.

## 3 Competitive Analysis

Since its introduction, Competitive Analysis [13, 18, 15] has been the most widely used method for evaluating online algorithms. In fact, the online search problem was first studied from the online algorithm perspective using Competitive Analysis by El-Yaniv et. al. [11]. Competitive Analysis evaluates an online algorithm in comparison to the optimal offline algorithm. Informally speaking, it considers the worst-case input which maximizes the ratio of the cost of the online algorithm for that input to the cost of the optimal offline algorithm on that same input. The maximum ratio achieved is called the competitive ratio.

**Definition 1** An online search algorithm $\mathcal{A}$ is *strictly c-competitive* if for all finite input sequences $I$,

$$OPT(I) \leq c \cdot \mathcal{A}(I).$$

The *competitive ratio* of algorithm $\mathcal{A}$ is $\inf\{c \mid \mathcal{A}$ is $c$-competitive$\}$. □

Denote the competitive ratio of an online algorithm $\mathcal{A}$ by $c_\mathcal{A}$. If $c_\mathcal{A} > c_\mathcal{B}$, $\mathcal{B}$ is better than $\mathcal{A}$ according to Competitive Analysis and we denote this by $\mathcal{A} \prec_c \mathcal{B}$.

In [11], El-Yaniv formulated the Reservation Price Policy algorithm and applied it to both deterministic and randomized search for real-valued prices. It was shown that the *reservation price* $p^* = \sqrt{Mm}$ is the optimal price for any deterministic online algorithm according to Competitive Analysis and using this price, the competitive ratio is $\sqrt{M/m}$. A very similar result holds for integer-valued prices.

**Theorem 1** According to Competitive Analysis, $\mathcal{R}_p \prec_c \mathcal{R}_q$, $\mathcal{R}_p \equiv_c \mathcal{R}_q$ and $\mathcal{R}_q \prec_c \mathcal{R}_p$ if and only if $Mm > p(q-1)$, $Mm = p(q-1)$ and $Mm < p(q-1)$, respectively.

**Proof** In any price sequence for an RPP algorithm $\mathcal{R}_p$, we can observe two cases: (i) all the prices are less than $p$, in which case the performance ratio, offline to online, will be at most $\frac{p-1}{m}$ with equality when there is a price $p-1$ and the last price is $m$; and (ii) at least one price is greater than or equal to $p$, in which case the offline to online performance ratio would be at most $\frac{M}{p}$ with equality when the first price greater than or equal to $p$ is exactly $p$ and there is another price $M$ somewhere later. So, the competitive ratio of $\mathcal{R}_p$ will be $c_{\mathcal{R}_p} = max(\frac{p-1}{m}, \frac{M}{p})$. It is easy to observe that $c_{\mathcal{R}_p} > c_{\mathcal{R}_q}$ if and only if $\frac{M}{p} > \frac{q-1}{m}$ since $\frac{p-1}{m} < \frac{q-1}{m}$ and $\frac{M}{p} > \frac{M}{q}$. This argument proves that $\mathcal{R}_p \prec_c \mathcal{R}_q$ if and only if $Mm > p(q-1)$. Similarly we can conclude the other two results. $\square$

**Corollary 1** Let $s = \left\lceil \sqrt{Mm} \right\rceil$. According to Competitive Analysis, the best RPP algorithm is $\mathcal{R}_s$.

**Proof** Assume that $p < s$. The comparison between $\mathcal{R}_s$ and $\mathcal{R}_p$ gives $p\left(\left\lceil \sqrt{Mm} \right\rceil - 1\right) < Mm$. Thus, $\forall p < s, \mathcal{R}_p \prec_c \mathcal{R}_s$.

Now, assume that $q > s$. Then the comparison between $\mathcal{R}_q$ and $\mathcal{R}_s$ gives $\left\lceil \sqrt{Mm} \right\rceil (q-1) \geq Mm$. Thus, $\forall q > s, \mathcal{R}_q \prec_c \mathcal{R}_s$.

$\square$

**Theorem 2** According to Competitive Analysis, $\mathcal{R}_p^2 \prec_c \mathcal{R}_p$ and $\mathcal{R}_p^2 \equiv_c \mathcal{R}_p$ if and only if $p > m$ and $p = m$, respectively.

**Proof** From the proof of Theorem 1, we know that the competitive ratio of $\mathcal{R}_p$ is $c_{\mathcal{R}_p} = \max\left(\frac{p-1}{m}, \frac{M}{p}\right)$. For the competitive ratio of $\mathcal{R}_p^2$, we consider a price sequence with only one $M$ followed by $n-1$ occurrences of $m$. Clearly the competitive ratio is $M/m$, and it is the maximum ratio that can be obtained by any algorithm. So, $c_{\mathcal{R}_p^2} \geq c_{\mathcal{R}_p}$, and equality holds if and only if $p = m$. $\square$

## 4 Bijective Analysis

In the Bijective Analysis model [1, 2], we construct a bijection on the set of possible input sequences. In this bijection, we aim to pair input sequences for online algorithms $\mathcal{A}$ and $\mathcal{B}$ in such a way that the cost of $\mathcal{A}$ on every sequence $I$ is no more than the cost of $\mathcal{B}$ on the image of $I$, or vice versa, to show that the algorithms are comparable. We present a version of the definition from [1] which is suitable for profit maximization problems such as online search.

**Definition 2** We say that an online search algorithm $\mathcal{A}$ is no better than an online search algorithm $\mathcal{B}$ according to *Bijective Analysis* if there exists an integer $n_0 \geq 1$ such that for each $n \geq n_0$, there is a bijection $b : \mathcal{I}_n \leftrightarrow \mathcal{I}_n$ satisfying $\mathcal{A}(I) \leq \mathcal{B}(b(I))$ for each $I \in \mathcal{I}_n$. We denote this by $\mathcal{A} \preceq_b \mathcal{B}$. $\square$

**Theorem 3** According to Bijective Analysis, $\mathcal{R}_p \prec_b \mathcal{R}_q$, if $p = m$ and $m < q \leq M$. Otherwise, $\mathcal{R}_p$ and $\mathcal{R}_q$ are incomparable.

**Proof** Consider the sequences with $m < p$. Note that with $\mathcal{R}_p$, $m$ will be chosen as output if and only if it is the last price of the sequence and all the preceding prices are smaller than $p$. As there are $p - m$ such prices and, not counting the last price, there are $n - 1$ prices in the sequence, the number of possible sequence with $m$ as output is $(p-m)^{n-1}$. With the same reasoning, for algorithm $\mathcal{R}_p$, each price in the range of $m$ to $p - 1$ will be the output for $(p - m)^{n-1}$ sequences.

For any prices in the range of $p$ to $M$, algorithm $\mathcal{R}_p$ chooses this price as output at its first occurrence in the price sequence if no price greater than or equal to $p$ has occurred before it. So all the preceding prices before this first occurrence should be smaller than $p$ (specifically in the range of $m$ to $p - 1$) and all of the next prices can be any value. For example, the number of sequences where price $p$ comes in the 3rd place in the sequence as well as taken as output will be $(p - m)^2 N^{n-3}$. So the number of sequences with any value $k$ in the range from $m$ to $M$ as output is

$$N_{p,k} = \begin{cases} (p-m)^{n-1}, & \text{for } m \leq k < p \\ \sum_{i=1}^{n}(p-m)^{i-1}N^{n-i}, & \text{for } p \leq k \leq M \end{cases} \quad (1)$$

where $N_{p,k}$ is the number of sequences which give output $k$ if the reservation price is $p$.

Now we explore the different conditions on the values of $p$ and $q$ for comparison of $\mathcal{R}_p$ and $\mathcal{R}_q$. Recall the assumption throughout the paper that $q > p$.

**Case $p > m$:** From Eq. (1), we can derive the fact that when $p > m$, the number of the sequences with lowest output for algorithm $\mathcal{R}_q$ ($N_{q,m}$) is greater than that of algorithm $\mathcal{R}_p$ ($N_{p,m}$) since $(q-m)^{n-1} > (p-m)^{n-1}$. Thus, we cannot have any bijective mapping $b: \mathcal{I}_n \leftrightarrow \mathcal{I}_n$ that shows $\mathcal{R}_p(I) \leq \mathcal{R}_q(b(I))$ for every $I \in \mathcal{I}_n$. On the other hand, it is also the case that the number of sequences with highest output ($M$) for algorithm $\mathcal{R}_q$ is greater than that of algorithm $\mathcal{R}_p$ since $N_{q,M} > N_{p,M}$. So $\mathcal{R}_p(I) \geq \mathcal{R}_q(b(I))$ is also not possible for every $I \in \mathcal{I}_n$. Thus for this case, $\mathcal{R}_p$ and $\mathcal{R}_q$ are incomparable according to the Bijective Analysis.

**Case $p = m$:** For algorithm $\mathcal{R}_m$, since the first price will be accepted, each price will be the output for exactly $N^{n-1}$ sequences. We can derive the number of sequences with specific output for algorithm $\mathcal{R}_q$ using Eq. (1). In this case, each price in the range from $m$ to $q-1$ will emerge as output in $(q-m)^{n-1}$ sequences and the number of sequences with output in the range from $q$ to $M$ will be $N^{n-1}+(q-m)N^{n-2}+(q-m)^2N^{n-3}+\cdots+(q-m)^{n-1}$. Clearly, here we can construct a bijective mapping $b: \mathcal{I}_n \leftrightarrow \mathcal{I}_n$ where each sequence with output $k < q$ of algorithm $\mathcal{R}_m$ is mapped to sequences with the same output for algorithm $\mathcal{R}_q$. Let $E_m$ denote the number of excess sequences with output $k < q$ of $\mathcal{R}_m$ which cannot be mapped in the above way. We map each sequence with output $k \geq q$ of algorithm $\mathcal{R}_m$ to sequences with the same output in algorithm $\mathcal{R}_q$. Let $E_q$ denote the number of excess sequences with output $k \geq q$ of $\mathcal{R}_q$ which can not be mapped in above way. Clearly, $E_m = E_q$ because the total number of sequences are the same in both algorithms. Note that, for all of these $E_m$ sequences, we can construct a mapping such that $\mathcal{R}_m(I) < \mathcal{R}_q(b(I))$. This mapping shows that $\mathcal{R}_m(I) \leq \mathcal{R}_q(b(I))$ for each $I \in \mathcal{I}_n$, but $\mathcal{R}_m(I) \geq \mathcal{R}_q(b(I))$ does not hold for all $I \in \mathcal{I}_n$. This shows that $\mathcal{R}_m$ and $\mathcal{R}_q$ are comparable according to Bijective Analysis and $\mathcal{R}_m \prec_b \mathcal{R}_q$. □

**Theorem 4** According the Bijective Analysis, $\mathcal{R}_p^2 \prec_b \mathcal{R}_p$ if $p > m$, and $\mathcal{R}_p^2 \equiv_b \mathcal{R}_p$ when $p = m$.

**Proof** Let $\hat{N}_{p,k}$ denote the number of sequences which give output $k$ for algorithm $\mathcal{R}_p^2$ if the reservation price is $p$. As in the case of $\mathcal{R}_p$, $\mathcal{R}_p^2$ will choose $m$ as the output if it is the last price of the sequence and all the preceding prices are smaller than than $p$. From the proof of the previous theorem, we know that there are exactly $(p-m)^{n-1}$ such sequences. In addition to those sequences, $m$ will also be the output if the preceding $n-1$ prices have exactly one price greater than or equal to $p$. The above reasoning is valid for each price in the range of $m$ to $p-1$. So,

$$\hat{N}_{p,k} = (p-m)^{n-1} + (p-m)^{n-2}(n-1)(M-p+1), \text{ for } m \leq k < p \quad (2)$$

For any price in the range of $p$ to $M$, algorithm $\mathcal{R}_p^2$ chooses this price as output if in the price sequence there is exactly one price greater than or equal to $p$ which occurs before it, or it is the last price and no other price $p$ or larger occurred earlier. So the number of sequences with any output $k$ in the range from $p$ to $M$ is

$$\hat{N}_{p,k} = \sum_{i=1}^{n}(p-m)^{i-2}(i-1)(M-p+1)N^{n-i}+(p-m)^{n-1} \text{ for } p \leq k \leq M \quad (3)$$

From Eq. (1) and (2), it is evident that if $p > m$ and $n > 1$, $\hat{N}_{p,k} > N_{p,k}$ for $m \leq k < p$. As the total number sequences is fixed and neither Eq. (1) nor Eq. (3) depend on $k$, $\hat{N}_{p,k} < N_{p,k}$ for $p \leq k < M$ when $p > m$. So, we can always build a bijective mapping $b : \mathcal{I}_n \leftrightarrow \mathcal{I}_n$ that has $\mathcal{R}_p^2(I) \leq \mathcal{R}_p(b(I))$ for every $I \in \mathcal{I}_n$, but the reverse mapping cannot be constructed if $p > m$. However, if $p = m$, then $\hat{N}_{p,k} = N_{p,k}$ for any $k$. So, according to Bijective Analysis, $\mathcal{R}_p^2 \prec_b \mathcal{R}_p$ if $p > m$ and $\mathcal{R}_p^2 \equiv_b \mathcal{R}_p$ when $p = m$. □

### 4.1 Real-Valued Price Interval

Note that, for real-valued instances of the problem, $N$ is not finite. The result of comparing the two algorithms using Bijective Analysis changes dramatically when the values of the prices are real numbers. Bijective Analysis cannot differentiate between algorithms when the number of sequences is uncountable.

**Theorem 5** $\mathcal{R}_p$ and $\mathcal{R}_q$ are equivalent according to Bijective Analysis if the prices are drawn from real space in $[m, M]$.

**Proof** As any closed or open interval in real space has the cardinality of the continuum, the cardinality of $[m, p]$ and $[m, q]$ will be same. Taking the Cartesian product of such sets preserves their cardinality. So the cardinality of the set of sequences with any output, $k$, will be same for both algorithms. Hence, we can find a bijective mapping between the sequences where each sequence is mapped to another with same output. This shows that in this situation, all reservation price algorithms are equivalent according to Bijective Analysis. □

The same problem clearly arises for other online problems with real-valued inputs.

## 5 Average Analysis

Using Bijective Analysis, algorithms would often be incomparable because it is impossible to find any mapping such that one algorithm dominates the other on all sequences and their images. However, in practical scenarios we often see that in spite of some exceptions, one algorithm performs much better than others for most of the cases. If we take the average performance of any algorithm, then we can overcome this type of problem. In [1], Average Analysis is defined with that aim and is here formulated in terms of online search.

**Definition 3** We say that an online search algorithm $\mathcal{A}$ is no better than an online search algorithm $\mathcal{B}$ according to *Average Analysis* if there exists an integer $n_0 \geq 1$ such that for each $n \geq n_0$, $\sum_{I \in \mathcal{I}_n} \mathcal{A}(I) \leq \sum_{I \in \mathcal{I}_n} \mathcal{B}(I)$. We denote this by $\mathcal{A} \preceq_a \mathcal{B}$. □

**Theorem 6** For all $n \geq \left\lfloor \frac{\log(N/(q-p))}{\log(N/(N-1))} \right\rfloor + 1$, $\sum_{I \in \mathcal{I}_n} \mathcal{R}_p(I) < \sum_{I \in \mathcal{I}_n} \mathcal{R}_q(I)$. Thus, according to Average Analysis, $\mathcal{R}_p \prec_a \mathcal{R}_q$.

**Proof** Let $S_{p,n}$ denote the summation $\sum_{I \in \mathcal{I}_n} \mathcal{R}_p(I)$. For the online search problem, we can derive the value of $S_{p,n}$ using Eq. (1) as follows:

$$
\begin{aligned}
S_{p,n} &= \sum_{i=m}^{p-1} i N_{p,i} + \sum_{i=p}^{M} i N_{p,i} \\
&= (p-m)^{n-1} \sum_{i=m}^{p-1} i + \left( \sum_{i=1}^{n} (p-m)^{i-1} N^{n-i} \right) \sum_{i=p}^{M} i \\
&= \frac{(p-m)^n (p+m-1)}{2} + \left( \sum_{i=1}^{n} (p-m)^{i-1} N^{n-i} \right) \frac{(M+p)(M-p+1)}{2} \quad (4) \\
&= \frac{(p-m)^n (p+m-1)}{2} + \left( \frac{N^n - (p-m)^n}{N - (p-m)} \right) \frac{(N+m+p-1)(N+m-p)}{2} \\
&= \frac{N^{n+1} + pN^n + mN^n - N^n - N(p-m)^n}{2}
\end{aligned}
$$
(5)

where we have used that $N = M - m + 1$.

To compare two algorithms, we essentially want to know if the difference between the corresponding two sums $(S_{q,n} - S_{p,n})$ is greater than zero for some $n_0 \geq 1$ and for each $n \geq n_0$. To derive the condition for $\mathcal{R}_p \prec_a \mathcal{R}_q$ using Derivation (5), we have

$$
\begin{aligned}
& S_{q,n} - S_{p,n} > 0 \\
\Leftrightarrow \quad & \frac{qN^n - pN^n - (q-m)^n N + (p-m)^n N}{2} > 0 \\
\Leftrightarrow \quad & N^{n-1} > \frac{(q-m)^n - (p-m)^n}{q-p}
\end{aligned}
$$

Now we have to find a $n_0 \geq 1$ such that for each $n \geq n_0$, the condition of Eq. (6) is satisfied. If we choose $n_0$ such that $N^{n_0-1} > \frac{(N-1)^{n_0}}{q-p}$, then the condition of Eq. (6) is fulfilled for $n_0$, as the maximum value of $q - m$ is $N - 1$ and $q > p$. So,

$$\begin{aligned}
& (n_0 - 1) \log N > n_0 \log(N - 1) - \log(q - p) \\
\Leftrightarrow\ & n_0(\log N - \log(N - 1)) > \log N - \log(q - p) \\
\Leftrightarrow\ & n_0 > \frac{\log(N/(q-p))}{\log(N/(N-1))}
\end{aligned} \tag{6}$$

Eq. (6) gives a value for $n_0$ of $\left\lfloor \frac{\log(N/(q-p))}{\log(N/(N-1))} \right\rfloor + 1$ for which for all $n \geq n_0$, $\sum_{I \in \mathcal{I}_n} \mathcal{R}_p(I) < \sum_{I \in \mathcal{I}_n} \mathcal{R}_q(I)$. This proves that $\mathcal{R}_p \prec_a \mathcal{R}_q$ according to Average Analysis. □

**Corollary 2** According to Average Analysis, the best RPP algorithm is $\mathcal{R}_M$.

**Proof** According to Theorem 6, for any two RPP algorithm $\mathcal{R}_p$ and $\mathcal{R}_q$ with $q > p$, $\mathcal{R}_p \prec_a \mathcal{R}_q$. Thus, the maximal possible reservation price is best. □

**Theorem 7** According to Average Analysis, $\mathcal{R}_p^2 \prec_a \mathcal{R}_p$ if $p > m$, and $\mathcal{R}_p^2 \equiv_a \mathcal{R}_p$ when $p = m$.

**Proof** To prove the relative performance of $\mathcal{R}_p^2$ and $\mathcal{R}_p$, we calculate the values of the expressions $\sum_{I \in \mathcal{I}_n} \mathcal{R}_p^2(I)$ and $\sum_{I \in \mathcal{I}_n} \mathcal{R}_p(I)$. To derive these values we use the expressions for $N_{p,k}$ and $\hat{N}_{p,k}$ from Eq. (1), (2), and (3) of the previous section. If $p = m$, then $N_{p,k} = \hat{N}_{p,k} = N^{n-1}$ for any value of $k$. Thus, if $p = m$, then $\sum_{I \in \mathcal{I}_n} \mathcal{R}_p^2(I) = \sum_{I \in \mathcal{I}_n} \mathcal{R}_p(I)$.

Now we consider the case where $p > m$. From Eq. (1), we know that the $N_{p,k}$ attain two distinct values, one in the range of $m \leq k < p$ and the other in $p \leq k \leq M$, i.e., $N_{p,m} = N_{p,m+1} = \ldots = N_{p,p-1}$ and $N_{p,p} = N_{p,p+1} = \ldots = N_{p,M}$. This is also true for $\hat{N}_{p,k}$ from Eq. (2) and (3). To prove $\mathcal{R}_p^2 \prec_a \mathcal{R}_p$ and consequently to show $\sum_{I \in \mathcal{I}_n} \mathcal{R}_p^2(I) < \sum_{I \in \mathcal{I}_n} \mathcal{R}_p(I)$, we develop following identity.

$$\sum_{k=m}^{M} N_{p,k} = \sum_{k=m}^{M} \hat{N}_{p,k} = N^n$$
$$\Leftrightarrow (p-m)N_{p,m} + (M-p+1)N_{p,M} = (p-m)\hat{N}_{p,m} + (M-p+1)\hat{N}_{p,M}$$
$$\Leftrightarrow (M-p+1)(N_{p,M} - \hat{N}_{p,M}) = (p-m)(\hat{N}_{p,m} - N_{p,m}) \quad (7)$$

Note that both $N_{p,M} - \hat{N}_{p,M}$ and $\hat{N}_{p,m} - N_{p,m}$ in the last equality are positive. Now we show that the difference between $\sum_{I \in \mathcal{I}_n} \mathcal{R}_p(I)$ and $\sum_{I \in \mathcal{I}_n} \mathcal{R}_p^2(I)$ is positive.

$$\sum_{I \in \mathcal{I}_n} \mathcal{R}_p(I) - \sum_{I \in \mathcal{I}_n} \mathcal{R}_p^2(I)$$
$$= \sum_{k=m}^{p-1} k N_{p,m} + \sum_{k=p}^{M} k N_{p,M} - \sum_{k=m}^{p-1} k \hat{N}_{p,m} - \sum_{k=p}^{M} k \hat{N}_{p,M}$$
$$= (N_{p,M} - \hat{N}_{p,M}) \sum_{k=p}^{M} k - (\hat{N}_{p,m} - N_{p,m}) \sum_{k=m}^{p-1} k$$
$$= (N_{p,M} - \hat{N}_{p,M}) \frac{(M-p+1)(M+p)}{2} - (\hat{N}_{p,m} - N_{p,m}) \frac{(p-m)(p+m-1)}{2}$$
$$= (N_{p,M} - \hat{N}_{p,M})(M-p+1)\left(\frac{M+p}{2} - \frac{p+m-1}{2}\right) \quad (8)$$
$$> 0 \quad (9)$$

where the second to last step follows from Eq. (7), and the last step from $M + p > p + m - 1$ and $N_{p,M} > \hat{N}_{p,M}$.

This shows that $\sum_{I \in \mathcal{I}_n} \mathcal{R}_p^2(I) < \sum_{I \in \mathcal{I}_n} \mathcal{R}_p(I)$ and accordingly proves that $\mathcal{R}_p^2 \prec_a \mathcal{R}_p$ if $p > m$, and $\mathcal{R}_p^2 \equiv_a \mathcal{R}_p$ when $p = m$. □

From the definition of Average Analysis, we know that it calculates the sum of outputs from all possible input sequences. For integral valued problems, the sum of the outputs and the expected output over a uniform distribution can be derived easily from each other, and either of them can be used to compare two algorithms. In contrast, in the case of real-valued problem, calculating the sum of the outputs is not possible as the number of sequences is infinite. However, if we know the distribution of the input prices in the sequences, then we can derive the expected output of a sequence. Here, we generalize Average Analysis to *Expected Analysis*. This generalization may prove useful for other online problems as well.

**Definition 4** We say that an online search algorithm $\mathcal{A}$ is no better than an online search algorithm $\mathcal{B}$ according to *Expected Analysis* if there exists an integer $n_0 \geq 1$ such that for each $n \geq n_0$, $E_{I \in \mathcal{I}_n}[\mathcal{A}(I)] \leq E_{I \in \mathcal{I}_n}[\mathcal{B}(I)]$. We denote this by $\mathcal{A} \preceq_e \mathcal{B}$. □

We denote the probability of the first price being from the range $[m, p)$ by $P_{m,p}$ and the probability of the first price being from the range $[p, M]$ by $P_{p,M}$. Additionally we denote the expected value of the prices less than $p$ by $E_{m,p}$ and the expected value of the prices greater than or equal to $p$ by $E_{p,M}$. We assume that the prices in an input sequence are independent and uniformly distributed over the range of $[m, M]$. If $n = 1$, then we can get the expected value of the output as $P_{p,M} E_{p,M} + P_{m,p} E_{m,p}$. Assume now that we are dealing with sequences of length two. Hence, with probability $P_{p,M}$, the algorithm $\mathcal{R}_p$ accepts the first price and with probability $P_{m,p}$ it will not. So, for $n = 2$, the expected value of the output will be $P_{p,M} E_{p,M} + P_{m,p} P_{p,M} E_{p,M} + P_{m,p} P_{m,p} E_{m,p}$. Inductively, the expected value of the output for a sequence of length $n$ can be calculated from that of sequences of length $n - 1$.

$$E_{I \in \mathcal{I}_n}[\mathcal{R}_p(I)] = P_{p,M} E_{p,M} \sum_{i=1}^{n} P_{m,p}^{i-1} + E_{m,p} P_{m,p}^n \qquad (10)$$

To analyze the performance of any RPP algorithm using Expected Analysis, we derive the values of the required probabilities and expectations. For the real-valued case, $P_{m,p} = \frac{p-m}{M-m}$, $P_{p,M} = \frac{M-p}{M-m}$, $E_{m,p} = \frac{p+m}{2}$, and $E_{p,M} = \frac{M+p}{2}$. For the integral case, $P_{m,p} = \frac{p-m}{M-m+1}$, $P_{p,M} = \frac{M-p+1}{M-m+1}$, $E_{m,p} = \frac{p+m-1}{2}$, and $E_{p,M} = \frac{M+p}{2}$. For the integral case of the algorithm $\mathcal{R}_p$, the above values give

$$E_{I \in \mathcal{I}_n}[\mathcal{R}_p(I)] =$$

$$\frac{(M-p+1)(M+p)}{2(M-m+1)} \sum_{i=1}^{n} \left(\frac{p-m}{M-m+1}\right)^{i-1} + \frac{p+m-1}{2} \left(\frac{p-m}{M-m+1}\right)^n$$

It is easily verifiable from the Eq. (4) that $E_{I \in \mathcal{I}_n}[\mathcal{R}_p(I)] = S_{p,n}/N^n$. Thus, Definition 3 and 4 produce the same result as stated in Theorem 6 for the integral case.

It is clear that neither Definition 3 nor Theorem 6 can be used in the real-valued case.

Intuitively, we can say that $\mathcal{R}_m$ always chooses the first price of any sequence, whereas $\mathcal{R}_M$ does not choose any of the first $n-1$ prices as $P_{M,M} = 0$ and it has to take the last price. As all the prices are identically distributed, the expected value of the first price and the last price are same. That makes $\mathcal{R}_m$ and $\mathcal{R}_M$ equivalent. For the rest of the cases, we denote the distance between $m$ and $M$ by $U$, i.e. $U = M - m$. This gives us the following.

**Proposition 1** *In case of real-valued online search*

$$E_{I \in \mathcal{I}_n}[\mathcal{R}_m(I)] = E_{I \in \mathcal{I}_n}[\mathcal{R}_M(I)] = \frac{m + M}{2}$$

Thus, according to Expected Analysis, $\mathcal{R}_m \equiv_e \mathcal{R}_M$.

**Theorem 8** *In case of real-valued online search, for all $n \geq \left\lfloor \frac{\log(U/(q-p))}{\log(U/(q-m))} \right\rfloor + 1$, if either $p > m$ or $q < M$, $E_{I \in \mathcal{I}_n}[\mathcal{R}_p(I)] < E_{I \in \mathcal{I}_n}[\mathcal{R}_q(I)]$. Thus, in this case, according to Expected Analysis, $\mathcal{R}_p \preceq_e \mathcal{R}_q$.*

**Proof** For the real-valued case, using Eq. (10), the expression $E_{I \in \mathcal{I}_n}[\mathcal{R}_p(I)]$ becomes

$$
\begin{aligned}
E_{I \in \mathcal{I}_n}[\mathcal{R}_p(I)] &= \frac{(M-p)(M+p)}{2U} \sum_{i=1}^{n} \left(\frac{p-m}{U}\right)^{i-1} + \frac{p+m}{2}\left(\frac{p-m}{U}\right)^n \\
&= \frac{(p-m)^n(p+m)}{2U^n} + \frac{(M+p)(M-p)}{2U^n} \sum_{i=1}^{n} (p-m)^{i-1} U^{n-i} \\
&= \frac{1}{2U^n}\left[(p-m)^n(p+m) + (M+p)(M-p)\frac{U^n - (p-m)^n}{U - (p-m)}\right] \\
&= \frac{1}{2U^n}[MU^n + pU^n - U(p-m)^n]
\end{aligned}
$$
(11)

To prove that $\mathcal{R}_p \prec_e \mathcal{R}_q$, it is sufficient to show that the difference between the corresponding two expectations is greater than zero for some $n_0 \geq 1$ and for each $n \geq n_0$. To derive this condition we use Eq. (11).

$$\begin{aligned}
&E_{I \in \mathcal{I}_n}[\mathcal{R}_q(I)] - E_{I \in \mathcal{I}_n}[\mathcal{R}_p(I)] > 0 \\
\Leftrightarrow \quad & \frac{qU^n - pU^n - (q-m)^n U + (p-m)^n U}{2U^n} > 0 \\
\Leftrightarrow \quad & U^{n-1} > \frac{(q-m)^n - (p-m)^n}{q-p}
\end{aligned} \tag{12}$$

Since $q < M$, Eq. (12) becomes similar to Eq. (6), with the only difference being that here $U = M - m$ in place of $N = M - m + 1$.

To get the value of $n_0$, we can follow the derivation of Eq. (6), concluding that in the case of real-valued problems, for all $n \geq \left\lfloor \frac{\log(U/(q-p))}{\log(U/(q-m))} \right\rfloor + 1$, $E_{I \in \mathcal{I}_n}[\mathcal{R}_p(I)] < E_{I \in \mathcal{I}_n}[\mathcal{R}_q(I)]$, if either $p > m$ or $q < M$. From the previous statement and Proposition 1, this proves that according to Expected Analysis, $\mathcal{R}_p \preceq_e \mathcal{R}_q$. □

## 6 Random Order Analysis

Kenyon [16] proposed another method for comparing the average behaviors of online algorithms by considering the expected result of a random ordering of an input sequence. In [16], Kenyon defines the random order ratio in the context of the bin packing problem which is a cost minimization problem.

**Definition 5** The *random order ratio* $RC(A)$ of an online bin packing algorithm $A$ is

$$RC(A) = \limsup_{OPT(I) \to \infty} \frac{E_\sigma[A(\sigma(I))]}{OPT(I)}$$

where the expectation is taken over all permutations of $I$. □

In other words, the random order ratio is the worst ratio obtained over all sequences $I$, comparing the expected value of an algorithm, $A$, to the value of $OPT$, the optimal offline algorithm, with respect to a uniform distribution on all permutations of $I$. An online algorithm $B$ is better than an online algorithm $A$ according to Random Order Analysis if $RC(A) > RC(B)$. We denote this by $A \prec_r B$. Since the value of $OPT(I)$ is bounded above by the constant $M$, the following definition, a maximization version of the definition of random order ratio in [14], is used here in place of the original definition.

$$RC(\mathcal{R}_p) = \limsup_{n \to \infty} \frac{OPT(I)}{E_\sigma[\mathcal{R}_p(\sigma(I))]} \qquad (13)$$

**Theorem 9** The random order ratio of the RPP algorithm $\mathcal{R}_p$ is $\max(\frac{M}{p}, \frac{p-1}{m})$ when $p > 1$ and $p > m$. Consequently, $\mathcal{R}_p \prec_r \mathcal{R}_q$ if and only if $Mm > p(q-1)$.

**Proof** Considering the random order ratio of $\mathcal{R}_p$, $OPT$ always get the highest price in the sequence as its output and $\mathcal{R}_p$ chooses the first price that is greater than or equal to $p$. There are two cases where the random order ratio could achieve the maximal value. First suppose the sequence has one price with value $M$ and all other prices are $p$. Then

$$E_\sigma[\mathcal{R}_p(\sigma(I))] = \frac{M(n-1)! + p(n-1)(n-1)!}{n!} \qquad (14)$$

Now substituting the expected value of the output of the algorithm $\mathcal{R}_p$ in Eq. (13), we can present the ratio as

$$\limsup_{n \to \infty} \frac{nM}{M + p(n-1)} = \frac{M}{p} \qquad (15)$$

The other case is when the sequence has one price with value $p-1$ and all other prices are equal to $m$. In this case, we can get a limit similar to Eq. (15) of $\frac{p-1}{m}$ when $p > 1$ and $p > m$. As we are seeking the maximum of these ratios, the random order ratio of $\mathcal{R}_p$ is the maximum of the two values, $\frac{M}{p}$ and $\frac{p-1}{m}$, when $p > 1$. This gives the same result as in the case of Competitive Analysis in Section 3. So, to prove the statement about the relative performance of $\mathcal{R}_p$ and $\mathcal{R}_q$, we can directly use the arguments provided in the proof of Theorem 1. □

**Corollary 3** Let $s = \lceil \sqrt{Mm} \rceil$. According to Random Order Analysis, the best RPP algorithm is $\mathcal{R}_s$.

**Proof** Since Random Order Analysis gives the same results as Competitive Analysis in comparing RPP algorithms, the best RPP algorithm is also the same (see Corollary 1). □

**Theorem 10** According to Random Order Analysis, $\mathcal{R}_p^2 \prec_r \mathcal{R}_p$ and $\mathcal{R}_p^2 \equiv_r \mathcal{R}_p$ if and only if $p > m$ and $p = m$, respectively.

**Proof** From the proof of Theorem 9, we know that the random order ratio of $\mathcal{R}_p$ is $RC(\mathcal{R}_p) = \max(\frac{p-1}{m}, \frac{M}{p})$. For the random order ratio of $\mathcal{R}_p^2$, we consider a price sequence with only one $M$ and $n-1$ occurrences of $m$. Clearly, $OPT$ always takes $M$, whereas $\mathcal{R}_p^2$ never accepts the first occurrence of $M$ unless it is the last price in the sequence. So,

$$E_\sigma[\mathcal{R}_p^2(\sigma(I))] = \frac{M(n-1)! + m(n-1)(n-1)!}{n!} \quad (16)$$

Now, substituting the expected value of the output of the algorithm $\mathcal{R}_p^2$ in Eq. (13), we can present the ratio as

$$RC(\mathcal{R}_p^2) = \limsup_{n \to \infty} \frac{nM}{M + m(n-1)} = \frac{M}{m} \quad (17)$$

This ratio is the maximum and worst ratio that can be obtained by any algorithm. So, $RC(\mathcal{R}_p) \geq RC(\mathcal{R}_p^2)$, and equality holds if and only if $p = m$. □

Note that the conditions are exactly the same as in the case of comparing $\mathcal{R}_p^2$ and $\mathcal{R}_p$ using Competitive Analysis in Section 3 and the proof of this theorem follows the proof of Theorem 2.

# 7 Relative Worst Order Analysis

Relative Worst Order Analysis [5] compares two online algorithms directly. It compares two algorithms on their worst orderings of sequences which have the same content, but possibly in different orders. The definition of this measure is somewhat more involved; see [6] for more intuition on the various elements. Here we use the definitions for the strict Relative Worst Order Analysis for profit maximization problems.

**Definition 6** Let $I$ be any input sequence, and let $n$ be the length of $I$. Let $A$ be any online search algorithm. Then

$$A_W(I) = \min_\sigma A(\sigma(I)).$$

□

**Definition 7** For any pair of algorithms $A$ and $B$, we define

$$c_l(A, B) = \sup\{c \mid \exists b : \forall I : A_W(I) \geq cB_W(I)\} \text{ and}$$
$$c_u(A, B) = \inf\{c \mid \exists b : \forall I : A_W(I) \leq cB_W(I)\}.$$

If $c_l(A, B) \geq 1$ or $c_u(A, B) \leq 1$, the algorithms are said to be *comparable* and the *strict relative worst order ratio* $WR_{A,B}$ of algorithm $A$ to algorithm $B$ is defined. Otherwise, $WR_{A,B}$ is undefined.

$$\text{If } c_l(A, B) \geq 1 \text{ then } WR_{A,B} = c_u(A, B), \text{ and}$$
$$\text{if } c_u(A, B) \leq 1 \text{ then } WR_{A,B} = c_l(A, B).$$

If $WR_{A,B} > 1$, algorithms $A$ and $B$ are said to be *comparable in $A$'s favor*. Similarly, if $WR_{A,B} < 1$, algorithms are said to be *comparable in $B$'s favor*. □

When two algorithms happen to be incomparable, Relative Worst Order Analysis can still be used to express their relative performance.

**Definition 8** If at least one of the ratios $c_u(A, B)$ and $c_u(B, A)$ is finite, the algorithms $A$ and $B$ are $(c_u(A, B), c_u(B, A))$-*related*. □

**Theorem 11** According to Relative Worst Order Analysis, $\mathcal{R}_q$ and $\mathcal{R}_p$ are $(\frac{M}{p}, \frac{q-1}{m})$-related. They are comparable in $\mathcal{R}_q$'s favor if $p = m$ and $q = m+1$.

**Proof** We compare two RPP algorithms, $\mathcal{R}_q$ and $\mathcal{R}_p$, with $p < q$, using strict Relative Worst Order Analysis. Notice that, unlike the other worst case analyses, we are not taking the worst ratio; rather, we are taking the worst profit from all the permutations of any sequence and then deriving the ratios. For the maximum value of the ratio of $\mathcal{R}_{q_w}(I)$ and $\mathcal{R}_{p_w}(I)$, we can construct a sequence $I$ with only one $p$ and one $M$ and all the other prices smaller than $q$. Among all the permutations of $I$, the worst output for $\mathcal{R}_q$ is $M$ and that of $\mathcal{R}_p$ is $p$. This gives the value of the upper bound $c_u(\mathcal{R}_q, \mathcal{R}_p)$ as $\frac{M}{p}$. For the lower bound, assume $I$ has only one $q - 1$ and one $m$ and all the other prices are smaller than $p$. Then, $\mathcal{R}_p$ takes $q - 1$ as its output on every permutation of $I$, but the worst output of $\mathcal{R}_q$ gives $m$. On this sequence, $\mathcal{R}_q$ performs worse than $\mathcal{R}_p$, and the ratio of the outputs of the two algorithms can never be lower than that. So,

$$c_l(\mathcal{R}_q, \mathcal{R}_p) = \frac{m}{q-1} \begin{cases} = 1, & \text{for } q = m+1 \text{ and } p = m \\ < 1, & \text{otherwise} \end{cases}$$
$$c_u(\mathcal{R}_q, \mathcal{R}_p) = \frac{M}{p} > 1$$

From the above expressions and the definitions of strict Relative Worst Order Analysis, we can see that $\mathcal{R}_q$ and $\mathcal{R}_p$ are comparable when $p = m$ and $q = m+1$. For all the other cases, they are incomparable. For this single feasible condition ($c_l(\mathcal{R}_q, \mathcal{R}_p) = 1$), we have $WR_{\mathcal{R}_q, \mathcal{R}_p} = \frac{M}{p} > 1$, and we can say that algorithms $\mathcal{R}_q$ and $\mathcal{R}_p$ are comparable in $\mathcal{R}_q$'s favor. Using Definition 8, since all the ratios are finite, $c_u(\mathcal{R}_p, \mathcal{R}_q)$ is $\frac{q-1}{m}$ and the algorithms $\mathcal{R}_q$ and $\mathcal{R}_p$ are $(\frac{M}{p}, \frac{q-1}{m})$-related. □

Note that this relatedness result gives the same conditions indicating which algorithm is better as Competitive and Random Order Analysis. Although the original definition of relatedness in Relative Worst Order Ratio does not tell explicitly which algorithm is better, we can get a strong indication about it from the next corollary.

**Corollary 4** Let $s = \lceil \sqrt{Mm} \rceil$. Then $\forall q > s$, if $\mathcal{R}_q$ and $\mathcal{R}_s$ are $(c, c')$-related then $c \leq c'$ and $\forall p < s$, if $\mathcal{R}_s$ and $\mathcal{R}_p$ are $(c, c')$-related then $c' > c$.

**Proof** By Theorem 11, $c_u(\mathcal{R}_q, \mathcal{R}_s) = \frac{M}{\lceil \sqrt{Mm} \rceil}$ and $c_u(\mathcal{R}_s, \mathcal{R}_q) = \frac{q-1}{m}$. As $q > s = \lceil \sqrt{Mm} \rceil$, $c_u(\mathcal{R}_q, \mathcal{R}_s) \leq c_u(\mathcal{R}_s, \mathcal{R}_q)$. On the other hand, $c_u(\mathcal{R}_s, \mathcal{R}_p) = \frac{M}{p}$ and $c_u(\mathcal{R}_p, \mathcal{R}_s) = \frac{\lceil \sqrt{Mm} \rceil - 1}{m}$. As $p < s = \lceil \sqrt{Mm} \rceil$, $c_u(\mathcal{R}_p, \mathcal{R}_s) < c_u(\mathcal{R}_s, \mathcal{R}_p)$. □

This corollary shows that whatever the value of $x$ ($x \neq s$), $c_u(\mathcal{R}_s, \mathcal{R}_x) \geq c_u(\mathcal{R}_x, \mathcal{R}_s)$. This could be defined as a weak form of optimality within a class of algorithms, and we will say that $\mathcal{R}_s$ *dominates* any other RPP algorithm.

**Theorem 12** According to Relative Worst Order Analysis, $\mathcal{R}_p$ and $\mathcal{R}_p^2$ are comparable in $\mathcal{R}_p$'s favor and $WR_{\mathcal{R}_p, \mathcal{R}_p^2} = \frac{M}{m}$.

**Proof** From the proofs of Theorems 2 and 10, we have already seen that the worst case performance ratio of $\mathcal{R}_p$ and $\mathcal{R}_p^2$ on the same sequence is $M/m$ and this is the highest possible value. So we can conclude that $c_u(\mathcal{R}_p, \mathcal{R}_p^2) = \frac{M}{m}$.

For deriving the lower bound of this ratio, notice that, for any sequence, on its worst permutation of that sequence, $\mathcal{R}_p$'s output will be at least as large as $\mathcal{R}_p^2$'s on its worst ordering of that sequence. We can prove this fact by taking the worst output of $\mathcal{R}_p$ and $\mathcal{R}_p^2$ over all the permutation of a sequence $I$. Let $x$ and $y$ denote these outputs, respectively. If $y < p$, then there is no price in $I$ smaller than $y$. So, the worst output of $\mathcal{R}_p$ must be greater than or equal to $y$, i.e., $x \geq y$. If $y \geq p$, it is the smallest price in $I$ greater than $p$, so again $x \geq y$. So, $c_l(\mathcal{R}_p, \mathcal{R}_p^2) = 1$. From Definition 7, we can conclude that according to Relative Worst Order Analysis, $\mathcal{R}_p$ and $\mathcal{R}_p^2$ are comparable in $\mathcal{R}_p$'s favor and $WR_{\mathcal{R}_p, \mathcal{R}_p^2} = \frac{M}{m}$. □

## 8 Relative Interval Analysis

Dorrigiv et. al. [10] proposed another analysis method, Relative Interval Analysis, in the context of paging. Relative Interval Analysis also compares two online algorithms directly, i.e., it does not use the optimal offline algorithm as the baseline of the comparison. It compares two algorithms on the basis of the rate of the outcomes over the length of the input sequence rather than their worst case behavior. Here we define this analysis for profit maximization problems for two algorithms $\mathcal{A}$ and $\mathcal{B}$, following [10].

**Definition 9** Let
$$Min_{\mathcal{A},\mathcal{B}}(n) = \min_{|I|=n} \{\mathcal{A}(I) - \mathcal{B}(I)\},$$

and
$$Max_{\mathcal{A},\mathcal{B}}(n) = \max_{|I|=n} \{\mathcal{A}(I) - \mathcal{B}(I)\}.$$

These functions are used to define the following two measures:

$$Min(\mathcal{A},\mathcal{B}) = \liminf_{n \to \infty} \frac{Min_{\mathcal{A},\mathcal{B}}(n)}{n} \text{ and } Max(\mathcal{A},\mathcal{B}) = \limsup_{n \to \infty} \frac{Max_{\mathcal{A},\mathcal{B}}(n)}{n}. \quad (18)$$

Note that $Min(\mathcal{A},\mathcal{B}) = -Max(\mathcal{B},\mathcal{A})$ and $Max(\mathcal{A},\mathcal{B}) = -Min(\mathcal{B},\mathcal{A})$. The *relative interval* of $\mathcal{A}$ and $\mathcal{B}$ is defined as

$$l(\mathcal{A},\mathcal{B}) = [Min(\mathcal{A},\mathcal{B}), Max(\mathcal{A},\mathcal{B})]. \quad (19)$$

If $Max(\mathcal{A},\mathcal{B}) > |Min(\mathcal{A},\mathcal{B})|$, then $\mathcal{A}$ is said to have better performance than $\mathcal{B}$ in this model. In particular, if $l(\mathcal{A},\mathcal{B}) = [0, \beta]$ for $\beta > 0$, then it is said that $\mathcal{A}$ dominates $\mathcal{B}$ since $Min(\mathcal{A},\mathcal{B}) = 0$ indicates that $\mathcal{A}$ is never worse than $\mathcal{B}$ and $Max(\mathcal{A},\mathcal{B}) > 0$ says that $\mathcal{A}$ is better at least for some case(s).
□

Now we turn to the comparison of the RPP algorithms $\mathcal{R}_p$ and $\mathcal{R}_q$. In this case, the output of the algorithms is not dependent on the length of the price sequence. So, the rate of the outputs over the length of the input sequence makes little sense; both $Min(\mathcal{A}, \mathcal{B})$ and $Max(\mathcal{A}, \mathcal{B})$ are zero for any algorithms. In online search, a more meaningful rate of output is over $N$, the number of possible prices. We assume that $m$ is a finite integer. We have modified Definition 9 accordingly and calculate the rate of output over $N$, using $N \to \infty$ to derive the limits of Eq. (18).

**Theorem 13** According to Relative Interval Analysis, $l(\mathcal{R}_q, \mathcal{R}_p) = [0, 1]$, if $q \in o(N)$.

**Proof** For the minimum value of $\mathcal{R}_q(I) - \mathcal{R}_p(I)$, we have any sequence of prices with all the prices smaller than $q$ where the first price is $q-1$ and the last price is $m$. In this case, $Min_{\mathcal{R}_q, \mathcal{R}_p}(N) = m - q + 1$. The maximum value of $\mathcal{R}_q(I) - \mathcal{R}_p(I)$ is $M - p$, when the first price is $p$ and the second price is $M$. To compare $\mathcal{R}_p$ and $\mathcal{R}_q$, we change Eq. (18) and we have the following expressions:

$$Min(\mathcal{R}_q, \mathcal{R}_p) = \liminf_{N \to \infty} \frac{m - q + 1}{N}, \text{ and } Max(\mathcal{R}_q, \mathcal{R}_p) = \limsup_{N \to \infty} \frac{M - p}{N}. \quad (20)$$

As $N = M - m + 1$ and $m < q \in o(N)$, these limits give $l(\mathcal{R}_q, \mathcal{R}_p) = [0, 1]$. This proves the theorem. Thus, $\mathcal{R}_q$ dominates $\mathcal{R}_p$. □

**Theorem 14** According to Relative Interval Analysis, $l(\mathcal{R}_p, \mathcal{R}_p^2) = [-1, 1]$, if $p \in o(N)$.

**Proof** For the minimum value of $\mathcal{R}_p(I) - \mathcal{R}_p^2(I)$, we use any sequence of prices starting with two prices in the order of $(p, M)$. In this case, $Min_{\mathcal{R}_p, \mathcal{R}_p^2}(N) = p - M$. The maximum value of $\mathcal{R}_p(I) - \mathcal{R}_p^2(I)$ is $M - m$, which occurs when the first price is $M$ and the other prices are $m$. The corresponding limits of Eq. (20) are:

$$Min(\mathcal{R}_p, \mathcal{R}_p^2) = \liminf_{N \to \infty} \frac{p - M}{N}, \text{ and } Max(\mathcal{R}_p, \mathcal{R}_p^2) = \limsup_{N \to \infty} \frac{M - m}{N}. \quad (21)$$

Clearly, the above two limits give $l(\mathcal{R}_p, \mathcal{R}_p^2) = [-1, 1]$ and according to Definition 9, we cannot conclude anything about the relative quality of these two algorithms. □

Given the finite nature of the online search problem, we also propose another modification of Relative Interval Analysis removing the limits from the definition.

**Definition 10** $Min_{\mathcal{A},\mathcal{B}}(n)$ and $Max_{\mathcal{A},\mathcal{B}}(n)$ are as in Definition 9.

These functions are used to define the following two measures:

$$Min(\mathcal{A},\mathcal{B}) = \inf\{Min_{\mathcal{A},\mathcal{B}}(n)\} \text{ and } Max(\mathcal{A},\mathcal{B}) = \sup\{Max_{\mathcal{A},\mathcal{B}}(n)\}. \quad (22)$$

Note that $Min(\mathcal{A},\mathcal{B}) = -Max(\mathcal{B},\mathcal{A})$ and $Max(\mathcal{A},\mathcal{B}) = -Min(\mathcal{B},\mathcal{A})$. The Finite Relative Interval of $\mathcal{A}$ and $\mathcal{B}$ is defined as

$$fl(\mathcal{A},\mathcal{B}) = [Min(\mathcal{A},\mathcal{B}), Max(\mathcal{A},\mathcal{B})]. \quad (23)$$

Relative performance and dominance with regards to $fl(\mathcal{A},\mathcal{B})$ are defined as for $l(\mathcal{A},\mathcal{B})$ from Definition 9. □

**Theorem 15** According to Finite Relative Interval Analysis, $fl(\mathcal{R}_q, \mathcal{R}_p) = [m - q + 1, M - p]$, and $fl(\mathcal{R}_p, \mathcal{R}_p^2) = [p - M, M - m]$.

**Proof** From the proof of Theorem 13, we know that for any value of $n$, $Min_{\mathcal{R}_q,\mathcal{R}_p}(n) = m - q + 1$ and $Max_{\mathcal{R}_q,\mathcal{R}_p}(n) = M - p$. That gives $fl(\mathcal{R}_q, \mathcal{R}_p) = [m - q + 1, M - p]$. From the proof of Theorem 14, we know that for any value of $n$, $Min_{\mathcal{R}_p,\mathcal{R}_p^2}(n) = p - M$ and $Max_{\mathcal{R}_p,\mathcal{R}_p^2}(n) = M - m$. That gives $fl(\mathcal{R}_p, \mathcal{R}_p^2) = [p - M, M - m]$. Together this shows that $\mathcal{R}_p$ has better performance than $\mathcal{R}_p^2$ if $p > m$. □

**Corollary 5** Let $s = \lceil \frac{M+m}{2} \rceil$. According to Finite Relative Interval Analysis, the best RPP algorithm is $\mathcal{R}_s$.

**Proof** Let $p < s$. To compare $\mathcal{R}_s$ and $\mathcal{R}_p$, we have $Min(\mathcal{R}_s, \mathcal{R}_p) = m - \lceil \frac{M+m}{2} \rceil + 1$ and $Max(\mathcal{R}_s, \mathcal{R}_p) = M - p > M - \lceil \frac{M+m}{2} \rceil$ as $p < \lceil \frac{M+m}{2} \rceil$. This shows that $\forall p < s$, $Max(\mathcal{R}_s, \mathcal{R}_p) > |Min(\mathcal{R}_s, \mathcal{R}_p)|$ and consequently we can say that $\mathcal{R}_s$ performs better than $\mathcal{R}_p$. Now assume $q > s$. Then a comparison between $\mathcal{R}_q$ and $\mathcal{R}_s$ gives $Min(\mathcal{R}_q, \mathcal{R}_s) = m - q + 1 < m - \lceil \frac{M+m}{2} \rceil + 1$ as $q > \lceil \frac{M+m}{2} \rceil$, and $Max(\mathcal{R}_q, \mathcal{R}_s) = M - s = M - \lceil \frac{M+m}{2} \rceil$. This inequality shows $\forall q > s$, $Max(\mathcal{R}_q, \mathcal{R}_s) \leq |Min(\mathcal{R}_q, \mathcal{R}_s)|$ and consequently we can say that $\mathcal{R}_s$ performs at least as well as $\mathcal{R}_q$. These two cases prove that $\mathcal{R}_s$ is a best RPP algorithm according to Finite Relative Interval Analysis. □

## 9 Max/Max Ratio Analysis

In [3], Ben-David et. al. defined the Max/Max ratio for cost minimization problems. The Max/Max ratio compares an algorithm's worst cost for any sequence of length $n$ to $OPT$'s worst cost for any sequence of length $n$. If we want to preserve this worst output ratio behavior for profit maximization problems, essentially the analysis must take the minimum profits and could be named Min/Min Ratio Analysis. Here we define the Min/Min ratio by modifying the definition of the Max/Max ratio.

**Definition 11** The Min/Min ratio of an online algorithm $A$, $w_M(A)$, is $M(OPT)/M(A)$, where

$$M(A) = \liminf_{n \to \infty} \min_{|I|=n} A(I)/n. \tag{24}$$

□

In the online search problem, for any RPP algorithm, the minimum output is $m$ for some sequence of length $n$. For example, the sequence of $n$ consecutive prices of value $m$ always has the output $m$. As $m$ is a finite value, the limit of Eq. (24) is zero for any algorithm. As online search problems are finite problem, Min/Min Ratio Analysis is not applicable in comparing online search algorithms.

However, we can modify the previous definition to make it suitable for finite problems.

**Definition 12** The Min/Min ratio of an online algorithm $A$, $w_M(A)$, is $M(OPT)/M(A)$, where

$$M(A) = \inf\{\min_{|I|=n} A(I)\}. \tag{25}$$

□

This definition gives $M(OPT)$ and $M(A)$ for any RPP algorithms $A$ the same value $m$ and consequently make the Min/Min ratio equal to 1. That makes every algorithm equivalent according to Min/Min Ratio Analysis.

## 10 Concluding Remarks

Our findings with regards to the study of performance measures are listed in the introduction and we will not repeat them here. Studying performance measures and disclosing their properties and differences from each other is work in progress. With this study, we have added Online Search to the collection of problems that have been investigated with a spectrum of measures. More online problem scenarios must be analyzed this broadly before strong conclusions concerning the different performance measures can be drawn.

One of the results of the analysis of online search, as explained in the paper, is that for an online player there are three choices for the optimal reservation prices, $\sqrt{mM}$, $\frac{m+M}{2}$, and $M$ depending on the different analysis methods, i.e., the geometric mean, the arithmetic mean and the maximum of the values $m$ and $M$. This clearly shows that the objectives of the different performance measures vary greatly, trying to limit poor performance in a proportional or additive sense, or focusing equally on all scenarios or placing emphasis on the limit. Thus, the different measures are tailored towards different degrees of risk aversion—cautiousness vs. aggressiveness. These observations complement the findings regarding greediness and laziness from [7]. We would like to encourage researchers to study (fundamentally) different online problems in a similar fashion to increase our understanding of the focus points for different performance measures.

## Acknowledgments

The authors are thankful to Leah Epstein, Asaf Levin and Alejandro López-Ortiz for earlier discussions concerning Bijective Analysis and problems related to real-valued inputs.